\documentclass{article}
\usepackage{bookstyle,bm,cite}
\usepackage{graphicx}
\begin{document}
\title{Random Matrices, the Ulam Problem, Directed Polymers \& Growth Models, and Sequence Matching}
\author{Satya N. Majumdar}
\address{
Laboratoire de Physique Th\'eorique et
Mod\`eles Statistiques (UMR 8626 du CNRS),
Universit\'e Paris-Sud, B\^at. 100, 91405 Orsay Cedex, France
}
\frontmatter
\maketitle
\mainmatter%
\section{Preamble}

In these lecture notes I will give a pedagogical introduction to some common aspects of $4$ different 
problems: (i) random matrices (ii) the longest increasing subsequence problem (also known as 
the Ulam problem) (iii) directed 
polymers in random medium
and growth models in $(1+1)$ dimensions and (iv) a problem on the alignment of a pair
of random sequences.
Each of these problems is almost entirely a sub-field by itself and here I will discuss only some specific
aspects of each of them. These $4$ problems have been studied almost independently for the
past few decades, but only over the last few years a common thread was found to
link all of them. In particular all of them share one common limiting probability distribution
known as the Tracy-Widom distribution that describes the asymptotic probability distribution
of the largest eigenvalue of a random matrix. I will mention here, without mathematical 
derivation, some of the
beautiful results discovered in the past few years. Then, I will consider two specific models
(a) a ballistic deposition growth model and (b) a model of sequence alignment known as the
Bernoulli matching model and discuss, in some detail, how one derives exactly the
Tracy-Widom law in these models. The emphasis of these lectures would be on how to
map one model to another. Some open problems will be discussed at the end.   

\section{Introduction}

\begin{figure}[t]
\includegraphics[width=.7\hsize]{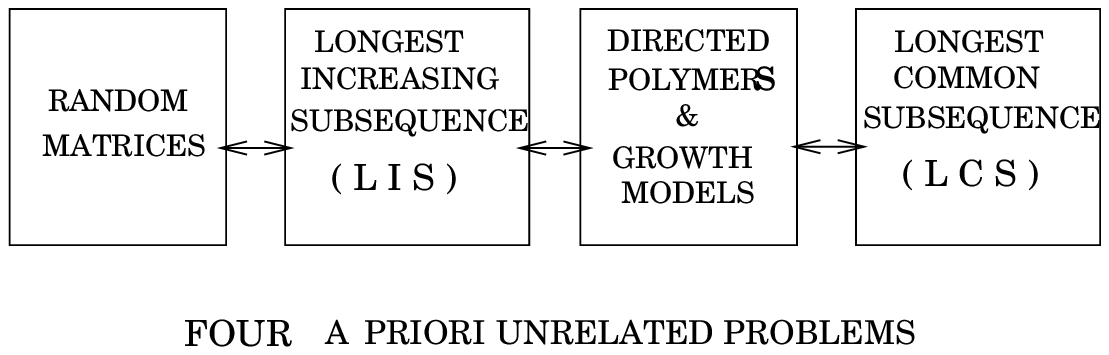}
\caption{All $4$ problems share the Tracy-Widom distribution.}
\label{fields}
\end{figure}

In these lectures I will discuss $4$ seemingly unrelated problems: (i) random matrices
(ii) the longest increasing subsequence (LIS) problem (also known as the Ulam
problem after its discoverer) (iii) directed polymers in random environment in $(1+1)$ dimensions
and related random growth models and (iv) the longest common subsequence (LCS)
problem arising in matching of a pair of random sequences (see Fig. \ref{fields}). These 4 problems
have been studied extensively, but almost independently, over the past few
decades.  
For example, random matrices have 
been extensively studied by
nuclear physicists, mathematicians and statisticians. The LIS problem
has been studied extensively by probabilists. The models of directed polymers in
random medium 
and the related growth models have been a very popular subject among
statistical physicists. Similarly, the LCS problem has been very popular
among biologists and computer scientists. Only, in the last $10$ years or so,
it became progressively evident that there are profound links between these
$4$ problems. All of them share one common probability distribution function
which is called the Tracy-Widom distribution.

This distribution was first discovered in the context of random matrices 
by Tracy and Widom~\cite{TW1}. They calculated exactly the probability distribution
of the {\em typical} fluctuations of the largest eigenvalue of a random matrix around
its mean. This distribution, suitably scaled, is known as the Tracy-Widom (TW)
distribution (see later for details). Later in 1999, in a landmark paper~\cite{BDJ},
Baik, Deift and Johansson (BDJ) showed that the same TW distribution
describes the scaled distributions of the length of the longest
increasing subsequence in the LIS problem. Immediately after, Johansson~\cite{J1},
Baik and Rains~\cite{BR1} showed that the same distribution also appears
in a class of directed polymer problems. Around the same time, Pr\"ahofer
and Spohn showed~\cite{PS} that the TW distribution also appears in
a class of random growth models known as the polynuclear growth (PNG) models.
Following this, it was discovered that the TW distribution 
also occurred 
in several other growth models, such as the `oriented digital boiling' model~\cite{GTW},
a ballistic deposition model~\cite{BD}, in PNG type of growth models
with varying initial conditions and in various geometries~\cite{IS,F1} and
also in the single-step growth model arising from the totally asymmetric exclusion process~\cite{S1}.
Also, a somewhat direct connection between the stochastic growth models
and the random matrix models via the so called `determinantal point processes' was found
in a series of work by Spohn and collaborators~\cite{Spohn}, which I will not discuss here
(see Ref. \cite{Spohn} for a recent review). 
Finally, the TW
distribution was also shown to appear in the LCS problem~\cite{MN}, which is also related
to these growth models.
Apart from these 4 problems that we will focus here, the TW distribution
has also appeared in many other problems, e.g., in the mesoscopic
fluctuations of excitation gaps in a dirty metal grain or a semiconductor quantum dot induced
by a nearby superconductor~\cite{meso}. 
The TW distribution also appears in problems related to finance~\cite{BBP}.  

The appearence of the TW distribution in so many different problems
is really interesting, suggesting an underlying universality 
that links all these different systems. The purpose of my lectures would
be to explore and elucidate the links between the 4 problems stated above. 
The literature on this subject is huge. I will not try to provide
any detailed derivation of the mathematical results here. Instead, 
I will state precisely the known results that we will need to use and put
more emphasis on how one maps one problem to the other. In particular,
I will discuss two problems in some detail and show how the TW distribution
appears in them. These two problems are: (i) a random growth model
in $(1+1)$ dimensions that we call the anisotropic ballistic deposition model
and (ii) a particular variant of the LCS problem known as the Bernoulli
matching (BM) model. In the former case, I will show how to the map
the ballistic deposition model to the LIS problem and subsequently use the BDJ
results. In the second case, I will show that the BM model can be mapped to 
a particular directed polymer model that was studied by Johansson.
The mappings are often geometric in nature, are nontrivial and serves
two purposes: (a) to elucidate how the TW distribution appears in
somewhat unrelated problems and (b) to derive exact analytical results in problems such
as the sequence matching models, where precise analytical results were
missing so far. 

The lecture notes are organized as follows. In Section 3, I will describe
some basic results of the random matrix theory and define the TW distribution
precisely. In Section 4, the LIS problem will be described along with
the main results of BDJ.
Section 5 contains a discussion of the directed polymer problems,
and in particular the main results of Johansson will be mentioned. In Section 5.1, I will describe
how one maps the anisotropic ballistic deposition model to the LIS problem.
Section 6 contains a discussion of the LCS problem. Finally, I will conclude
in Section 7 with a discussion and open problems.

\section{Random Matrices: the Tracy-Widom distribution for the largest eigenvalue}

Studies of the statistics of the eigenvalues of random matrices have a
long history going back to the seminal work of Wigner~\cite{Wigner}.
Since then, random matrices have found applications in multiple fields
including nuclear physics, quantum chaos, disordered systems, string
theory and number theory~\cite{Mehta}. Three classes of matrices with
Gaussian entries have played important roles~\cite{Mehta}: $(N\times
N)$ real symmetric (Gaussian Orthogonal Ensemble (GOE)), $(N\times N)$
complex Hermitian (Gaussian Unitary Ensemble (GUE)) and $(2N\times
2N)$ self-dual Hermitian matrices (Gaussian Symplectic Ensemble
(GSE)). For example, in GOE, one considers an $(N\times
N)$ real symmetric matrix $X$ whose elements $x_{ij}$'s are drawn
independently from a Gaussian distribution: $P(x_{ii})= \frac{1}{\sqrt{2\pi}}\,\exp[-x_{ii}^2/2]$
and $P(x_{ij}) = \frac{1}{\sqrt{\pi}}\,\exp[-x_{ij}^2]$ for $i<j$. Thus the
joint distribution of all the $N(N+1)/2$ independent elements is just the product
of the individual distributions and can be writen in a compact form as
$P[X]= A_N \exp[-{\rm tr}(X^2)/2]$, where $A_N$ is a normalization constant.
One can similarly write down the joint distribution for the other two ensembles~\cite{Mehta}.

One of the key results in the random matrix theory is due to Wigner who derived,
starting from the joint distribution of the matrix elements $P(X)$, a rather
compact expression for the
joint probability density function (PDF) of the eigenvalues of a random $(N\times
N)$ matrix from all ensembles~\cite{Wigner}
\begin{equation}
P(\lambda_1, \lambda_2,\dots, \lambda_N) = B_N \exp\left[-\frac{\beta}{2}\left(\sum_{i=1}^N\lambda_i^2
-\sum_{i\ne j}\ln(|\lambda_i-\lambda_j|)\right)\right],
\label{pdf}
\end{equation}
where $B_N$ normalizes the pdf and $\beta=1$, $2$ and $4$ correspond
respectively to the GOE, GUE and GSE. The joint law allows one to
interpret the eigenvalues as the positions of charged particles,
repelling each other via a $2$-d Coulomb potential (logarithmic);
they are confined on a $1$-d line and each is subject to an external harmonic
potential. The parameter $\beta$ that characterizes the type of
ensemble can be interpreted as the inverse temperature.

\begin{figure}
\includegraphics[width=.7\hsize]{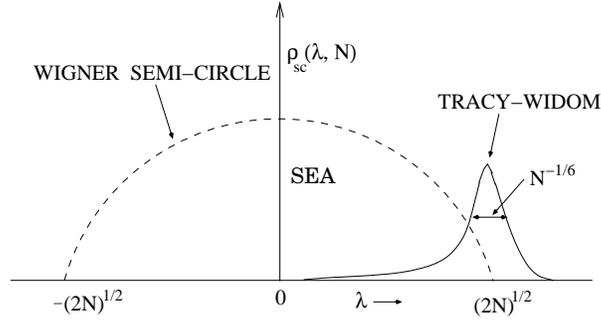}
\caption{The dashed line shows the semi-circular form of the
average density of states. The largest eigenvalue is centered around its mean $\sqrt{2N}$
and fluctuates over a scale of width $N^{-1/6}$. The probability of fluctuations
on this scale is described by the Tracy-Widom distribution (shown schematically).}
\label{figtw}
\end{figure}

Once the joint pdf is known explicitly, other statistical properties of a random matrix
can, in principle, be derived from this joint pdf. In practice, however
this is often a technically daunting task. For example, suppose we want to
compute the average density of states of the eigenvalues defined as
$\rho(\lambda,N)= \sum_{i=1}^N\langle
\delta(\lambda-\lambda_i)\rangle/N$, which counts the average number of
eigenvalues between $\lambda$ and $\lambda + d\lambda$ per unit length.
The angled bracket $\langle \rangle$ denotes an average over the joint pdf.
It then follows that $\rho(\lambda,N)$ is simply the marginal of the joint pdf,
i.e, we fix one of the eigenavlues (say the first one) at $\lambda$ and integrate the joint pdf
over the rest of the $(N-1)$ variables.
\begin{equation}
\rho(\lambda,N)=\frac{1}{N} \sum_{i=1}^N\langle
\delta(\lambda-\lambda_i)\rangle =\int_{-\infty}^{\infty}\prod_{i=2}^N d\lambda_i \, 
P(\lambda,\lambda_2,\dots, \lambda_N).
\label{marginal}
\end{equation}
Wigner was able to compute this marginal and this is one of the central results 
in the random matrix theory, known as the celebrated Wigner semi-circular law. For large $N$
and for any $\beta$,
\begin{equation}
\rho (\lambda,N) = \sqrt{\frac{2}{N\pi^2}}\,{\left[1 -\frac{\lambda^2}{2N}\right]}^{1/2}.
\label{wig1}
\end{equation}
Thus, on an average, the $N$ eigenvalues lie within a
finite interval $\left[-\sqrt{2N}, \sqrt{2N}\right]$, often referred
to as the Wigner `sea'. Within this sea, the average density of states
has a semi-circular form (see Fig. \ref{figtw}) that vanishes at the
two edges $-\sqrt{2N}$ and $\sqrt{2N}$. Note that since there are $N$
eigenvalues distributed over the interval $\left[-\sqrt{2N}, \sqrt{2N}\right]$, the
average spacing between adjacent eigenvalues scales as $N^{-1/2}$. 

From the semi-circular law, it is clear that the average of the maximum (or minimum) eigenvalue
is $\sqrt{2N}$ $\left(-\sqrt{2N}\right)$. However, for finite but large $N$, the maximum
eigenvalue fluctuates, around its mean $\sqrt{2N}$, from one sample to
another. A natural question is: what is the full probability distribution
of the largest eigenvalue $\lambda_{\rm max}$? Once again, this distribution
can, in principle, be computed from the joint pdf in Eq. (\ref{pdf}). To see
this, it is useful to consider the cumulative distribution of $\lambda_{\rm max}$.
Clearly, if $\lambda_{\rm max}\le t$, it necessarily means that all the eigenvalues
are less than or equal to $t$. Thus, 
\begin{equation}
{\rm Prob}\left[\lambda_{\rm max}\le t, N\right]= \int_{-\infty}^t \prod_{i=1}^N d\lambda_i \,
P(\lambda_1,\lambda_2,\dots, \lambda_N),
\label{max1}
\end{equation}
where the joint pdf is given in Eq. (\ref{pdf}).
In practice, however, carrying out this multiple integration in closed form is very difficult.
Relatively recently, Tracy and Widom~\cite{TW1} were 
able to find the limiting form of ${\rm Prob}\left[\lambda_{\rm 
max}\le t, 
N\right]$ for large $N$. They showed that the fluctuations of $\lambda_{\rm max}$ 
{\em typically} occur over a very narrow scale of 
width $\sim N^{-1/6}$ around its mean $\sqrt{2N}$ at the upper edge of the Wigner sea.
It is useful to note that this scale $\sim N^{-1/6}$ of typical fluctuations
of the largest eigenvalue is much bigger than the average spacing $\sim N^{-1/2}$
between adjacent eigenvalues in the limit of large $N$.

\begin{figure}
\includegraphics[width=.7\hsize]{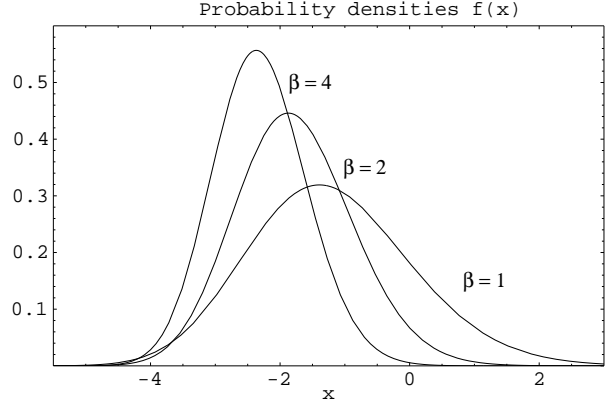}
\caption{The probability density function $f_{\beta}(x)$ plotted as a
function of $x$ for $\beta=1$, $2$ and $4$ (reproduced from Ref. ~\cite{TW1}).}
\label{fig:tracy}
\end{figure}

More precisely, Tracy and Widom showed~\cite{TW1} that asymptotically for
large $N$, the scaling variable $\xi=\sqrt{2}\,N^{1/6}\, \left[\lambda_{\rm
max}-\sqrt{2N}\right]$ has a limiting $N$-independent probability
distribution, ${\rm Prob}[\xi\le x]= F_{\beta}(x)$ whose form depends
on the value of the parameter $\beta=1$, $2$ and $4$ characterizing
respectively the GOE, GUE and GSE. The function $F_{\beta}(x)$ is called
the Tracy-Widom (TW) distribution function. The function $F_{\beta}(x)$,
computed as a solution of a nonlinear Painleve differential equation~\cite{TW1},
approaches to $1$ as $x\to \infty$ and decays rapidly to zero as $x\to
-\infty$. For example, for $\beta=2$, $F_2(x)$ has the following
tails~\cite{TW1},
\begin{eqnarray}
F_2(x) &\to & 1- O\left(\exp[-4x^{3/2}/3]\right)\quad\, {\rm as}\,\,\, x\to \infty
\nonumber \\
&\to & \exp[-|x|^3/12] \quad\, {\rm as}\,\,\, x\to -\infty.
\label{asymp1}
\end{eqnarray}
The probability density function $f_{\beta}(x)=dF_{\beta}/dx$ thus has highly
asymmetric tails. A graph of these functions for $\beta=1$, $2$ and $4$
is shown in Fig. \ref{fig:tracy}. 
A convenient way to express these typical fluctuations of $\lambda_{\rm max}$
around its mean $\sqrt{2N}$ is to write, for large $N$, 
\begin{equation}
\lambda_{\max} = \sqrt{2N} + \frac{N^{-1/6}}{\sqrt{2}}\, \chi
\label{tw2}
\end{equation}
where the random variable $\chi$ has the limiting $N$-independent distribution,
${\rm Prob}[\chi \le x] = F_{\beta}(x)$.
As mentioned in the introduction, amazingly this TW distribution function has since
emerged in a growing variety of seemingly unrelated problems, some of which I
will discuss in the next sections.
\vspace{0.4cm}

{\bf {Large Deviations of $\lambda_{\rm max}$:}} Before we end this section and proceed to the other 
problems, it is worth making 
the following remark. The Tracy-Widom distribution describes the probability of {\em typical and small}
fluctuations of $\lambda_{\rm max}$ over a very narrow region of width
$\sim O(N^{-1/6})$ around the mean $\langle \lambda_{\rm max}\rangle
\approx \sqrt{2N}$. A natural question is how to describe the
probability of {\em atypical and large} fluctuations of $\lambda_{max}$ around its
mean, say over a wider region of width $\sim O(N^{1/2})$? For example,
what is the probability that all the eigenvalues of a random matrix
are negative (or equivalently all are positive)?  This is the same as
the probability that $\lambda_{\rm max}\le 0$ (or equivalently
$\lambda_{\rm min}\ge 0$). Since $\langle \lambda_{\rm max}\rangle
\approx \sqrt{2N} $, this requires the computation of the probability
of an extremely rare event characterizing a large deviation of $\sim
-O(N^{1/2})$ to the left of the mean. 
This question naturally arises in any physical system where one
is interested in the statistics of stationary points of a random landscape.
For example, in disordered systems such as spin glasses one is interested in
the stationary points (metastable states) of the free energy landscape.
On the other hand, in structural glasses or supercooled liquids, one is
interested in the stationary points of the potential energy landscape.
In order to have a local minimum of the
random landscape one needs to ensure that the eigenvalues of the
associated Hessian matrix are all positive~\cite{CGG,Fyodorov}.
A similar question recently came up
in the context of random landscape models of anthropic principle
based string theory~\cite{Susskind,AE} as well as in quantum
cosmology~\cite{MH}.  Here one is interested in the statistical
properties of vacua associated with a random multifield potential,
e.g., how many minima are there in a random string landscape? 
These large deviations are also important in characterizing the large sample
to sample fluctuations of the excitation gap in quantum dots
connected to a superconductor~\cite{meso}.

The issue of large deviations of $\lambda_{\rm max}$ was addressed 
in Ref. \cite{J1} for a special class of matrices drawn
from the Laguerre ensemble that corresponds to the eigenvalues of product
matrices of the form $W=X^{\dagger}X$ where $X$ itself is a Gaussian
matrix (real or complex). Adopting similar methods as in 
Ref. \cite{J1}
one can prove that for Gaussian ensembles, 
the probability of {\em large} fluctuations to the left of the mean $\sqrt{2N}$
behaves for large $N$ as,
\begin{equation}
{\rm Prob}\left[\lambda_{\rm max}\le t, N\right] \sim \exp\left[-\beta
N^2 \Phi_{-}\left( \frac{\sqrt{2N}-t}{\sqrt{N}} \right) \right]
\label{ldf1}
\end{equation}
where $t\sim O(N^{1/2})\le \sqrt{2N}$ is located deep inside the
Wigner sea and $\Phi_{-}(y)$ is a certain {\em left} large deviation function. 
On the other hand, for {\em large} fluctuations to the right of the mean $\sqrt{2N}$,  
\begin{equation}
1-{\rm Prob}\left[\lambda_{\rm max}\le t, N\right] \sim \exp\left[-\beta
N \Phi_{+}\left( \frac{t-\sqrt{2N}}{\sqrt{N}} \right) \right]
\label{ldf2}
\end{equation}
for $t\sim O(N^{1/2})\ge \sqrt{2N}$ located outside the Wigner sea to its right
and $\Phi_{+}(y)$ is the {\em right} large deviation function. 
The problem then is to evaluate explicitly the left and the right large deviation
functions $\Phi_{\mp}(y)$ explicitly. 
While, for the Laguerre ensemble, an explicit
expression of $\Phi_{+}(y)$ was obtained in Ref. \cite{J1} and 
that of $\Phi_(y)$ recently in Ref. \cite{VMB}, similar expressions
for the Gaussian ensemble were missing so far.

Indeed, to calculate the probability
that all eigenvalues are negative (or positive) for Gaussian matrices, we need an explicit expression
of $\Phi_{-}(y)$ for the Gaussian ensemble. This is because, the probability that all
eigenvalues are negative is precisely the probability that $\lambda_{\rm max}\le 0$,
and hence, from Eq. (\ref{ldf1})
\begin{equation}
{\rm Prob}\left[\lambda_{\rm max}\le 0, N\right]\sim \exp[-\beta N^2 \Phi_{-}(\sqrt{2})].
\label{exp1}
\end{equation}
The coefficient $\theta= \beta \Phi_{-}(\sqrt{2})$ of the $N^2$ term inside
the exponential term in Eq. (\ref{exp1}) is of interest in string theory,
and in Ref. \cite{AE}, the authors provided an approximate estimate (for $\beta=1$) of  
$\theta \approx 1/4$, along with numerical simulations. 
Recently, in collaboration with D.S. Dean,
we were able to compute exactly an explicit expression~\cite{DM} for
the full {\em left} large deviation function $\Phi_{-}(y)$. 
I will not provide the derivation here, but the calculation of 
{\em large} deviations turns out to be somewhat simpler~\cite{DM} than the calculation of the {\em small} 
deviations `a la TW. One simply has to minimize the effective free energy
of a Coulomb gas using the method of steepest descents and then analyze the
resulting saddle point equation (which is an integral equation)~\cite{DM}.
This technique is quite useful, as it can be applied to other problems
as well, such as the calculation of the average number of stationary points
for a Gaussian random fields with $N$ components in the large $N$ limit~\cite{BrayDean,FSW}
and also the large deviation function associated with the largest eigenvalue
of other types of matrices, such as the Wishart matrices~\cite{VMB}.  
In terms of the variable $z=y-\sqrt{2}$, the {\em left} large deviation
function has the following 
explicit expression~\cite{DM}
\begin{eqnarray}
\Phi_{-}(y=z+\sqrt{2})& =& -\frac{1}{8}(3+2 \ln 2) + \frac{1}{216}\left[ 72z^2 -2z^4 
(30z + 2z^3) \sqrt{6+z^2} \right.\nonumber \\
&+& \left. 27\left( 3 + \ln(1296) - 4 \ln\left(-z +
\sqrt{6 +z^2}\right) \right) \right].
\label{ldfl}
\end{eqnarray}
In particular, the constant $\theta$ is given exactly by
\begin{equation}
\theta = \beta\, \Phi(\sqrt{2})= \beta\, \frac{\ln 3}{4} = (0.274653\dots )\,\beta.
\label{theta}
\end{equation} 

Another interesting point about the left large deviation function $\Phi_{-}(y)$ is the following.
It describes the probability of large $\sim O(\sqrt{N})$ fluctuations to the left of the mean, i.e.,
when $y=(\sqrt{2N}-\lambda_{\rm max})/\sqrt{N} \sim O(1)$. Now, if we take the $y\to 0$ limit,
then $\Phi_{-}(y)$ should describe the {\em small} fluctuations to the left of the mean $\sqrt{2N}$.
In other words, we expect to recover the left tail of the TW distribution by taking the $y\to 0$
limit in the left large deviation function. Indeed, as $y\to 0$, one finds from Eq. (\ref{ldfl}),
that $\Phi_{-}(y) \approx y^3/{6\sqrt{2}}$. Putting this expression back in Eq. (\ref{ldf1})
one gets
\begin{equation}
{\rm Prob}[\lambda_{\rm max}\le t, N]\approx \exp\left[-\frac{\beta}{24}\big|\sqrt{2}\,
N^{1/6}\,(t-\sqrt{2N})\big|^3\right]
\label{asymp3}
\end{equation}
Given that $\chi= \sqrt{2}\,
N^{1/6}\,\left(t-\sqrt{2N}\right)$ is the Tracy-Widom scaling variable, we find that the result
in Eq. (\ref{asymp3}) matches exactly with the left
tail of the Tracy-Widom distribution for all $\beta$.
For example, for $\beta=2$ one can easily verify this by comparing Eqs. (\ref{asymp3})
and (\ref{asymp1}).
This approach not only serves as a useful check that one has obtained the correct
large deviation function $\Phi_{-}(y)$, but also provides an alternative and simpler way 
to derive the asymptotics of the left tail of the TW distribution.
A similar expression for the right large deviation function $\Phi_+(y)$ for the
Gaussian ensemble is still missing and its computation remains an open problem.

Although the Tracy-Widom distribution was originally derived as the limiting distribution
of the largest eigenvalue of matrices whose elements are drawn from Gaussian distributions,
it is now believed that the same limiting distribution also holds for matrices drawn
from a larger class of ensembles, e.g., when the entries are independent
and identically distributed random variables drawn from an arbitrary distribution
with all moments finite~\cite{Sosh,BBP1}.
Recently, Biroli, Bouchaud and Potters ~\cite{BBP} extended this result to
power-law ensembles, where each entry of a random matrix is drawn independently
from a power-law distribution~\cite{CB,Burda}. 
They showed that
as long as the fourth moment of this power-law distribution is finite, the suitably
scaled $\lambda_{\rm max}$ is again TW distributed, but when the fourth moment is
infinite, $\lambda_{\rm max}$ has Fr\'echet fluctuations~\cite{BBP}. It would be interesting
to compute the probability of {\em large} deviations of $\lambda_{\rm max}$ 
for this power-law ensemble, as in the Gaussian case mentioned above. For example,
what is the probability that all the eigenavlues of such random matrices (drawn
from the power-law ensemble) are negative (or positive), i.e. $\lambda_{\rm max}\le 0$?
This is an open question.

\section{The Longest Common Subsequence Problem (or the Ulam Problem)}

The longest common subsequence (LIS) problem was first stated by Ulam~\cite{Ulam} in 1961, hence
it is also called the Ulam's problem. Since then, a lot of research, mostly by
probabilists, has been done on this problem (for a brief history of the problem, see the
introduction in Ref. \cite{BDJ}). The problem can be stated very simply as follows.
Consider a set of $N$ distinct integers $\{1,2,3,\dots, N\}$. Consider all 
$N!$ possible permutations of 
this sequence. For any given
permutation, let us find all possible increasing subsequences (terms of a
subsequence need not necessarily be consecutive elements) and from them find
out the longest one. For example, take $N=10$ and consider a particular
permutation $\{8, 2, 7, \underbar 1, \underbar 3, \underbar 4, 10, \underbar 6,
\underbar 9, 5\}$. From this sequence, one can form several increasing
subsequences such as $\{8,10\}$, $\{2,3,4,10\}$, $\{1,3,4,10\}$ etc. The
longest one of all such subsequences is either $\{1,3,4,6,9\}$ as shown by the
underscores or $\{2,3,4,6,9\}$. The length $l_N$ of the LIS
(in our example $l_N=5$) is a random
variable as it varies from one permutation to another. In the Ulam problem one
considers all the $N!$ permutations to be equally likely. Given this uniform
measure over the space of permutations, what is the statistics of the random
variable $l_N$?

Ulam found numerically that the average length $\langle
l_N\rangle$ behaves asymptotically $\langle l_N\rangle\sim c \sqrt{N}$ for
large $N$. Later this result was established rigorously by Hammersley
\cite{Hammersley} and the constant $c=2$ was found by Vershik and Kerov
\cite{VK}. Recently, in a seminal paper, Baik, Deift and Johansson (BDJ)
\cite{BDJ} derived the full distribution of $l_N$ for large $N$. In particular,
they showed that asymptotically for large $N$
\begin{equation}
l_N \to 2\sqrt {N} + N^{1/6} \chi
\label{lis1}
\end{equation}
where the
random variable $\chi$ has a limiting $N$-independent distribution,
\begin{equation}
{\rm Prob}(\chi\leq x) = F_2(x)
\label{gue}
\end{equation}
where $F_2(x)$ is precisely 
the TW distribution for the largest eigenvalue of a random matrix
drawn from the GUE ($\beta=2$), as defined in Section 3.
Note that the power of $N$ in the correction term in Eq. (\ref{lis1}) is ${+1/6}$
as opposed to the asymptotic law in Eq. (\ref{tw2}) where the power of $N$ in the correction term
is $-1/6$. This means that while for random matrices of size $(N\times N)$, the typical
fluctuation of $\lambda_{\rm max}$ around its mean value $\sqrt{2N}$ {\em decreases} with
$N$ as $N^{-1/6}$ as $N\to \infty$ (i.e., the distribution gets narrower ans narrower
around the mean as $N$ increases), the opposite happens in the Ulam problem: the
typical fluctuation in $l_N$ around its mean $2\sqrt{N}$ {\em increases} as $N^{1/6}$ 
with increasing $N$, i.e., the distribution around the mean gets broader and broader
with increasing $N$.

BDJ also showed that
when the sequence length $N$ itself is a random variable drawn from a
Poisson distribution with mean $\langle N\rangle =\lambda$, the length of the LIS converges for
large $\lambda$ to
\begin{equation}
l_{\lambda}\to 2\sqrt{\lambda} + {\lambda}^{1/6} \chi,
\label{bdj1}
\end{equation}
where $\chi$ has the Tracy-Widom distribution $F_2(x)$. The fixed $N$ and the fixed
$\lambda$ ensembles are like the canonical and the grand canonical ensembles in
statistical mechanics. The
BDJ results led to an avalanche of subsequent mathematical works \cite{AD}.
\begin{figure}[t]
\includegraphics[width=.7\hsize]{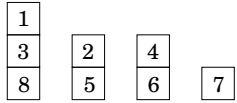}
\caption{The construction of piles according to the patience sorting game. The number
of piles corresponding to the sequence 
$\{8,3,5,1,2,6,4,7\}$ is $4$, which is also the length of the LIS of this sequence.}
\label{psheap}
\end{figure}

I will not provide here the derivation of the BDJ results, but I will assume this
result to be known and use it later for other problems. As we will see later, in
many problems such as in several growth models, the stratgey is to map those models
into the LIS problem and subsequently use the BDJ results. In these mappings, typically
the height of a growing surface in the $(1+1)$ dimensional growth models gets mapped to
the length of the LIS, i.e., schematically, $H \to l_N$. Subsequently, using the BDJ
results for the distribution of $l_N$, one shows that the height in growth models
is distributed accoriding to the Tracy-Widom law. I will show explicitly how this
strategy works for one specific ballistic deposition model in Section 5.1.
But to understand the mapping, we need to know one additional fact about the LIS, which I 
discuss below.

Suppose we are given a specific permutation of $N$ integers.
What is a simple algorithm to find the length of the LIS of this permuation?
The most famous algorithm goes by the name of Robinson-Schensted-Knuth (RSK)
algorithm~\cite{RSK}, which makes a correspondence between the permutation
and a Young tableaux, and has played a very important role in the development
of the LIS problem. But let me not discuss this 
here, the reader can find a nice readable account in Ref. \cite{AD}. Instead, I will
discuss another related algorithm known as the `patience-sorting' algorithm which
will be more useful for our purposes. This algorithm was developed first by Mallows~\cite{Mallows}
who showed its connection to the Young tableaux. I will discuss here the version that was
discussed recently by Aldous and Diaconis~\cite{AD}. This algorithm is best explained 
in terms of an example. Let us take $N=8$ and consider a specific permuation,
say $\{8,3,5,1,2,6,4,7\}$. The `patience sorting' is a greedy algorithm
that will easily find the length of the LIS of this sequence. It is like
a simple card game of `patience'. This game
goes as follows: start forming piles with the numbers in the permuted sequence
starting with the first element which is $8$ in our example. So, the number 8
forms the base of the first pile (see Fig. \ref{psheap}). The next element, if less than 8, goes on
top of 8. If not, it forms the base of a new pile. One follows a greedy
algorithm: for any new element of the sequence, check all the top numbers on
the existing piles starting from the first pile and if the new number is less
than the top number of an already existing pile, it goes on top of that pile.
If the new number is larger than all the top numbers of the existing piles,
this new number forms the base of a new pile. Thus in our example, we form $4$
distinct piles: $[\{8,3,1\}, \{5,2\}, \{6,4\}, \{7\}]$. Thus the number of piles 
is $4$. On the other hand, for this particular example, it is easy to check
that there are $3$ LIS's namely, $\{3,5,6,7\}$, $\{1,2,6,7\}$ and $\{1,2,4,7\}$, all of the same
length $l=4$. So, we see that the length of the LIS is $4$, same as the number of
piles in the patience sorting game. But this is not an accident. One can
easily prove~\cite{AD} that for any given permutation of $N$ integers, the length of the
LIS $l_N$ is exactly the same as the number of piles in the corresponding `patience sorting'
algorithm. We will see later that this fact does indeed play a crucial role in our mapping
of growth models to the LIS problem.

\section{Directed Polymers and Growth Models}

The problem of directed polymers in random medium has been an active area 
of research in statistical physics for the past three decades.
Apart from the fact that it is a simple `toy' model of disordered systems,
the directed polymer problem has important links to a wide variety
of other problems in physics, such as interface fluctuations and pinning~\cite{HH},
growing interface models of the Kardar-Parisi-Zhang (KPZ)
variety~\cite{KPZ}, randomly forced Burger's equation in fluid dynamics~\cite{FNS},
spin glasses~\cite{DS1,Mezard,FH}, 
and also to a single-particle quantum mechanics problem in a time-dependent random 
potential~\cite{Kardar}. There are many interesting issues associated 
with the directed polymer problem, such as the phase-transition at a finite
temperature in $(d+1)$-dimensional directer polymer when $d>2$~\cite{IS1}, the nature
of the low temperature phase~\cite{Mezard,FH}, the nature of the tranverse fluctuations~\cite{KZ,HH}
etc. The literature on the subject is huge (for a review see Ref. \cite{HZ}).

\begin{figure}[t]
\includegraphics[width=.7\hsize]{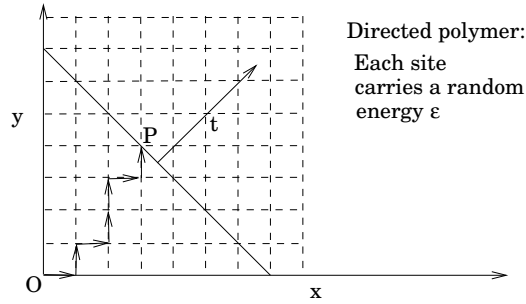}
\caption{Directed polymer in $(1+1)$ dimensions with random site energies.}
\label{dp}
\end{figure}

Here we will focus simply at zero-temperature and a lattice version of the directed polymer
problem. This version can be stated as in Fig. \ref{dp}.
Consider a square lattice with $O$ denoting the origin.
On each site with coordinates $(i,j)$ of this lattice, there is a random energy 
$\epsilon_{i,j}$, drawn 
independently
from site to site, but from the identical distribution $\rho(\epsilon)$. For simplicity, we
will consider that $\epsilon_{i,j}$'s are all negative, i.e., $\rho(\epsilon)$ has support
only over $\epsilon\in [0,-\infty]$. The energy variables $\epsilon_{i,j}$'s are quenched
random variables. 

We are interested here only in directed walks for simplicity.
Consider all possible directed walk configurations (a walk that can move only 
north or eastward as shown in Fig. \ref{dp}) that start from the origin $O$ and end up
at a fixed point, say $P$ with co-ordinates $(x,y)$. An example of such a walk
is shown in Fig. \ref{dp}.
The total energy $E(W)$ of any given walk $W$ from $O$ to $P$ is just the sum of site energies along the path 
$W$,
$E(W)= \sum_{i\in W} \epsilon_i$. Thus, for fixed $O$ and $P$ (the endpoints), the energy of a
path varies from one path to another (all having
the same endpoints $O$ and $P$). The path having the minimum energy (optimal path) among these will
correspond to the ground state configuration, i.e., the polymer will prefer to choose
this optimal path at zero temperature. Let $E_0(x,y)$ denote this minimum energy amongst
all directed paths that start at $O$ and finish at $P:(x,y)$. Now, this minimum energy
$E_0(x,y)$ is, of course, a random variable since it fluctuates from one configuration 
of quenched disorder to another. One is interested in the statistics of $E_0(x,y)$ for
a given fixed $(x,y)$. For example, what is the probability distribution of $E_0(x,y)$
given that $\epsilon_{x,y}$'s are independent and identically distributed random variables each 
drawn from $\rho(\epsilon)$? 

Mathematically, one can write an `evolution' equation or recursion relation for the variable $E_0(x,y)$. 
Indeed, the path that ends up at say $(x,y)$, must have visited either the site $(x-1,y)$
or the site $(x,y-1)$ at the previous step. Then clearly,
\begin{equation}
E_0(x,y) = {\rm min}\left[E_0(x-1,y), E_0(x,y-1)\right] + \epsilon_{x,y}
\label{dpr1}
\end{equation}
where $\epsilon_{x,y}$ denotes the random energy associated with the site $(x,y)$. 
Alternately, we can define $H(x,y)=-E_0(x,y)$ which are all positive variables that
satisfy the recursion relation
\begin{equation}
H(x,y) = {\rm max}\left[H(x-1,y), H(x,y-1)\right] + \xi_{x,y}
\label{dpr2}
\end{equation}
where $\xi_{x,y}=-\epsilon_{x,y}$ are positive random variables. The recursion 
relation in Eq. (\ref{dpr2}) is non-linear and hence is difficult to find the
distribution of $H(x,y)$, knowing the distribution of the $\xi_{x,y}$'s.
Note that, by interpreting $t=x+y$ as a time-like variable, and denoting 
by $i$ the transverse coordinate at a fixed $t$, this recursion
relation can also be interpreted as a stochastic evolution equation,
\begin{equation}
H(i,t) = {\rm max}\left[H(i+1,t-1), H(i-1,t-1)\right] + \xi_{i,t}
\label{dpr3}
\end{equation}
where the site energy $\xi_{i,t}$ can now be interpreted as a stochastic noise.
In this interpretation, one can think of the directed polymer as a growing
model of $(1+1)$ dimensional interface where $H(i,t)$ denotes the height of the interface
at the site $i$ of a one dimensional lattice at time $t$. Only, in this version, the
length of one dimensional lattice or the substrate keeps increasing linearly with 
time $t$. In this respect, it corresponds to a special version of a polynuclear
growth model where growth occurs on top of a single droplet whose linear size
keeps increasing uniformly with time.
There are, of course, several other variations of this simple directed
polymer model~\cite{HZ}. For example, one can consider a version
where the random energies are associated with bonds, rather than the sites.
Similarly, one can consider a finite temperature version of the model.
In the corresponding analogy to the interface model, at finite temperature, the free energy
(as opposed to the ground state energy) of the polymer corresponds to the
height variable of the interface. This is most easily seen in the continuum formulation
of the model by writing down the partition function as a path integral
and then showing directly that $H=\ln Z$ satisfies the KPZ equation~\cite{HHF}. 

A lot is known about the first and the second moment of $H(x,y)$ (or alternatively
for $H(i,t)$ in the height language)
and the associated universality properties~\cite{Mezard,FH,KMH}. For example, from simple
extensivity properties, one would expect that average ground state energy
of the path will increase linearly with the size (number of steps $t$) of the path.
In terms of height, this means $\langle H(i,t)\rangle \to v(i) t$ for large $t$
where $v(i)$ is velocity of the interface at site $i$ of the one dimensional
lattice~\cite{KH}. Also, the standard deviation of height, 
say of $H(x,x)$ (along the diagonal),
is known to grow universally, for large $x$ as $x^{1/3}$~\cite{HZ}. For the interface, this means
that the typical height fluctuation grows as $t^{1/3}$ for large $t$, a result
that is known from the KPZ problem in $1$-dimension (via a mapping to the noisy
Burgers equation). 
However, much less was known about the full distribution
of $H(x,y)$, till only recently.

Johansson~\cite{J1} was able to derive the full asymptotic distribution of $H(x,y)$
evolving via Eq. (\ref{dpr2}) for a specific disorder distribution, where the noise
$\xi_{x,y}$'s in Eq. (\ref{dpr2}) are i.i.d variables taking nonnegative integer
values according to the distribution: ${\rm Prob}(\xi_{x,y}=k)= (1-p)\, p^k$ for $k=0,1,2,\dots$,
where $0\le p\le 1$ is a fraction.
Interestingly, exactly the same recursion relation as in Eq. (\ref{dpr2}) and also
with the same disorder distribution as in Johansson's model
also appeared independently around the same time in an anisotropic directed percolation
problem studied by Rajesh and Dhar~\cite{RD}, a problem to which we will come back
later when we discuss the sequence matching problem. The authors in Ref.~\cite{RD} were able
to compute exactly the first moment, but Johansson computed the full asymptotic
distribution. He showed that for large $x$ and $y$~\cite{J1} 
\begin{eqnarray}
H(x,y) &\to& \frac{2\sqrt{pxy}+p(x+y)}{q}+ \nonumber \\
       &+&   \frac{(pxy)^{1/6}}{q}\,\left[(1+p)+\sqrt{\frac{p}{xy}}\,(x+y)\right]^{2/3}
       \, \chi 
\label{j1}
\end{eqnarray}
where $q=1-p$, $\chi$ is a random variable with the Tracy-Widom distribution, ${\rm Prob}(\chi\le x)=F_2(x)$
as in Eq. (\ref{gue}). If one sets $x=y=t/2$, then for the growing droplet interpretation, it would
mean that the height $H(i=0,t)$ has a mean that grows linearly with $t$ and a standard deviation
that grows as $t^{1/3}$ and when properly centered and scaled, the distribution of $H(0,t)$
tends to the GUE Tracy-Widom distribution. Around the same time, Pr\"ahofer and Spohn derived
a similar result for a class of PNG models~\cite{PS}. Moreover, they were able to show that not just the 
$F_2(x)$,
but other Tracy-Widom distributions such as the $F_1(x)$ (corresponding to the GOE ensemble)
also arises in the PNG model when one starts from different initial conditions~\cite{PS}. 

\subsection{Exact Height Distribution in A Ballistic Deposition Model}

In this subsection, we will show explicitly how one can derive the exact height distribution
in a specific $(1+1)$ dimensional growth model and show that it has a limiting Tracy-Widom
distribution. This example will illustrate explicitly how one maps a growth model
to the LIS problem~\cite{BD}. A similar mapping was used by Pr\"ahofer and Spohn 
for the PNG model~\cite{PS}. But before we illustrate the mapping, it is useful 
to remark (i) why one studies such growth models and (ii) what does this mapping
and subsequent calculation of the height distribution achieve?

The answer to these two questions are as follows. We know that growth processes are
ubiquitous in nature. The past few decades have seen
extensive research on a wide variety of both discrete and contiuous growth models
\cite{Meakin,KS,HZ}. A large class of these growth models in $(1+1)$ dimensions 
such as the Eden model
\cite{Eden}, restricted solid on solid (RSOS) models \cite{RSOS}, directed
polymers as mentioned before~\cite{HZ}, polynuclear growth models (PNG) \cite{PNG} and ballistic
deposition models (BD)~\cite{BaD} are believed to belong to the same
universality class as that of the Kardar-Parisi-Zhang (KPZ) equation describing the
growth of interface fluctuations \cite{KPZ}. This universality is, however,
somewhat restricted in the sense that it refers only to the width or the second
moment of the height fluctuations characterized by two independent exponents
(the growth exponent $\beta$ and the dynamical exponent $z$) and the associated
scaling function. Moreover, even this restricted universality is established
mostly numerically. Only in very few special discrete models in $(1+1)$ dimensions, the
exponents $\beta=1/3$ and $z=3/2$ can be computed exactly via the Bethe ansatz
technique \cite{Bethe}. A natural and important question is whether this
universality can be extended beyond the second moment of height fluctuations.
For example, is the full distribution of the height fluctuations (suitably
scaled) universal, i.e. is the same for different growth models belonging to
the KPZ class? Moreover, the KPZ-type equations are usually attributed to
models with small gradients in the height profile and the question whether the
models with large gradients (such as the BD models) belong to the KPZ universality class is still
open. The connection between the discrete BD models and the continuum KPZ equation
has recently been elucidated upon \cite{KS1}.

To test whether this more stringent test of universality (going beyond the second moment) of the full
distribution is true or not,
one needs to calculate the full height distribution in different models which are known
to belong to the KPZ universality class as far as only the second moment is concerned.
In fact, as mentioned earlier, Pr\"ahofer and Spohn were able to calculate the asymptotic height 
distribution in a class of PNG models and showed that it has the Tracy-Widom distribution~\cite{PS}.
Similarly, we mentioneed earlier that Johansson~\cite{J1} established rigorously that 
the height distribution,
in a specific version of the directed polymer model, is of the Tracy-Widom form.
Subsequently, there have been several other works~\cite{GTW} recently, including the ballistic deposition
model~\cite{BD} that we will discuss below, that showed that indeed 
all these $(1+1)$ dimensional growth models share the same common scaled height distribution   
(Tracy-Widom), thus putting the universality on a much stronger footing going beyond just the
second moment.  

\begin{figure}
\includegraphics[width=.7\hsize]{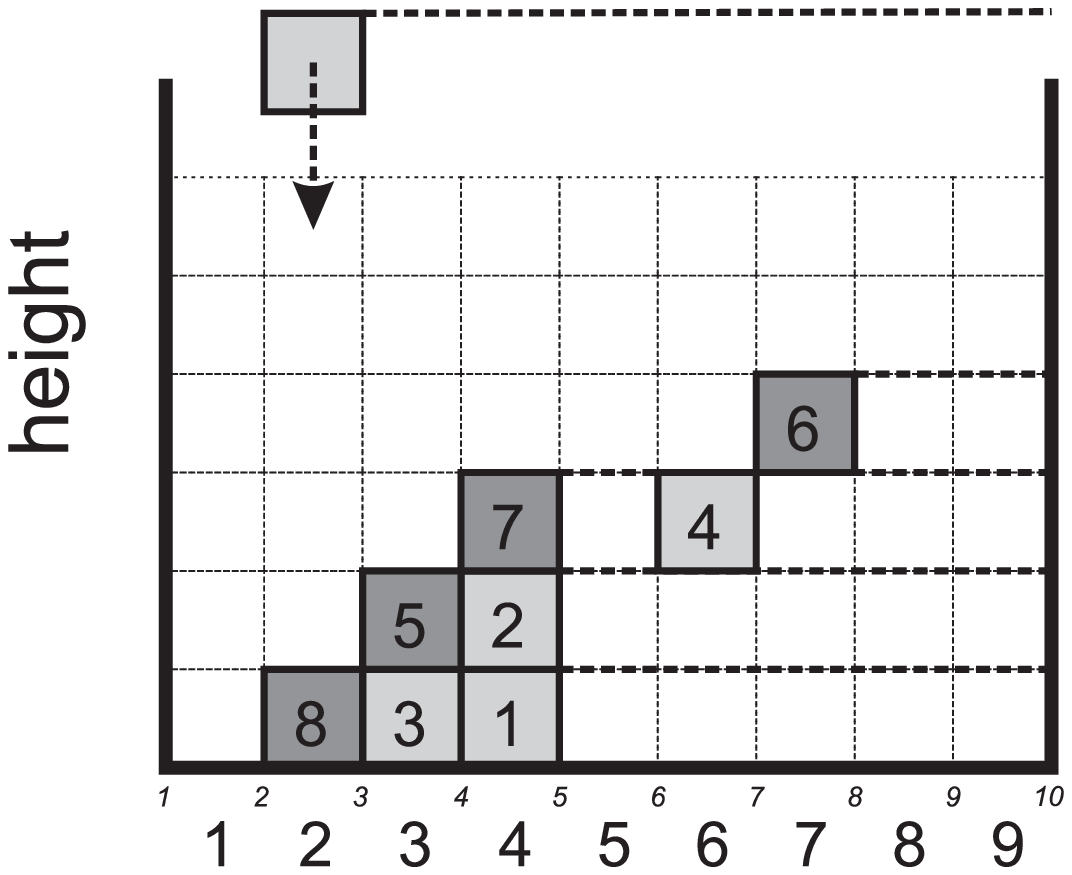}
\caption{Growth of a heap with asymmetric long-range interaction. The numbers
inside cells show the times at which the blocks are added to the heap.}
\label{fig:1}
\end{figure}

We now focus on a specific ballistic deposition model. Ballistic deposition models typically
try to mimic columnar growth that occur in many natural systems and have been studied
extensively in the past with a variety of microscopic rules~\cite{Krug2,BaD}, though an exact calculation
of the height distribution remained elusive in any of these microscopic models. In collaboration
with S. Nechaev, we found a particular ballistic deposition model which can be explicitly mapped
to the LIS problem and hence the full asymptotic height distribution can be computed
exactly~\cite{BD}. 
In our $(1+1)$-D (here $D$ stands for `dimensional') BD model columnar growth occurs sequentially on a linear 
substrate
consisting of $L$ columns with free boundary conditions. The time $t$ is
discrete and is increased by $1$ with every deposition event. We first consider
the flat initial condition, i.e., an empty substrate at $t=0$. Other initial
conditions will be treated later. At any stage of the growth, a column (say the
$k$-th column) is chosen at random with probability $p=\frac{1}{L}$ and a
"brick" is deposited there which increases  the height of this column by one
unit, $H_k\to H_k+1$. Once this "brick" is deposited, it screens all the sites
at the same level in all the columns to its right from future deposition, i.e.
the heights at all the columns to the right of the $k$-th column must be
strictly greater than or equal to $H_k+1$ at all subsequent times. For example,
in Fig. \ref{fig:1}, the first brick (denoted by 1) gets deposited at $t=1$ in
the 4-th column and it immediately screens all the sites to its right. Then the
second brick (denoted by 2) gets deposited at $t=2$ again in the same 4-th
column whose height now becomes 2 and thus the heights of all the columns to
the right of the 4-th column must be $\ge 2$ at all subsequent times and so on.
Formally such growth is implemented by the following update rule. If the $k$-th site
is chosen at time $t$ for deposition, then
\begin{equation}
H_k(t+1)={\rm max}\{H_k(t), H_{k-1}(t), \dots, H_1(t)\}+1.
\label{update1}
\end{equation}
The model is anisotropic and evidently even the average height profile $\langle
H_k(t) \rangle$ depends nontrivially on both the column number $k$ and time
$t$. Our goal is to compute the asymptotic height distribution $P_k(H,t)$ for
large $t$.

It is easy to find the height distribution $P_1(H, t)$ of the first column,
since the height there does not depend on any other column. At any stage, the
height in the first column either increases by one unit with probability
$p=\frac{1}{L}$ (if this column is selected for deposit) or stays the same with
probability $1-p$. Thus $P_1(H,t)$ is simply the binomial distribution,
$P_1(H,t)={t\choose H}p^h(1-p)^{t-H}$ with $H\leq t$. The average height of the
first column thus increases as $\langle H_1(t)\rangle=pt$ for all $t$ and its
variance is given by $\sigma_1^2(t)= tp(1-p)$. While the first column is thus
trivial, the dynamics of heights in other columns is nontrivial due to the
right-handed infinite range interactions between the columns. For
convenience, we subsequently measure the height of any other column with respect to the
first one. Namely, by height $h_k(t)$ we mean the height difference between the
$(k+1)$-th column and the first one, $h_k(t)=H_{k+1}(t)-H_1(t)$, so that
$h_0(t)=0$ for all $t$.

\begin{figure}
\includegraphics[width=.7\hsize]{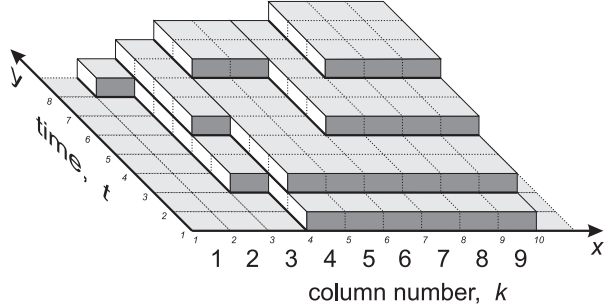}
\caption{$(2+1)$ dimensional "terraces" corresponding to the growth of a heap
in Fig. \ref{fig:1}}
\label{fig:2}
\end{figure}

To make progress for columns $k>0$, we first consider a
(2+1)-D construction of the heap as shown in Fig. \ref{fig:2}, by adding an extra
dimension indicating the time $t$. In Fig. \ref{fig:2}, the $x$ axis denotes the
column number, the $y$ axis stands for the time $t$ and the $z$ axis is the
height $h$. In this figure, every time a new block is added, it "wets" all the
sites at the same level to its "east" (along the $x$ axis) and to its "north"
(along the time axis). Here "wetting" means "screening" from
further deposition at those sites at the same level. This $(2+1)$-D system of
"terraces" is in one-to-one correspondence with the $(1+1)$-D heap in
Fig. \ref{fig:1}. This construction is reminiscent of the 3D anisotropic
directed percolation (ADP) problem studied by Rajesh and Dhar \cite{RD}. Note however,
that unlike the ADP problem, in our case each row labelled by $t$ can contain
only one deposition event.

\begin{figure}
\includegraphics[width=.7\hsize]{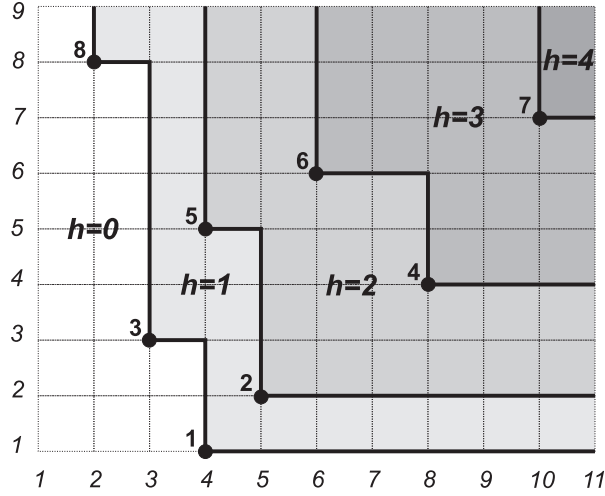}
\caption{The directed lines are the level lines separating adjacent terraces
with height diffrence $1$ in Fig. 2, projected onto the $(x,y)$ plane and
shifted by $(x,y)\to (x+h(x,y),y)$ to avoid partial overlap. The black dots
denote the deposition events. The numbers next to the dots denote the times of
those deposition events.}
\label{fig:3}
\end{figure}

The next step is to consider the projection onto the 2D $(x,y)$-plane of the
level lines separating  the adjacent terraces whose heights differ by $1$. In
this projection, some of the level lines may overlap partially on the plane.
To avoid the overlap for better visual purposes, we make a shift
$(x,y)\to (x+h(x,y),y)$ and represent these shifted directed lines on the 2D
plane in Fig. \ref{fig:3}.
The black dots in Fig. \ref{fig:3} denote the points
where the deposition events took place and the integer next to a dot denotes
the time of this event. Note that each row in Fig. \ref{fig:3} contains a single
black dot, i.e., only one deposition per unit of time can occur. In
Fig. \ref{fig:3}, there are 8 such events whose deposition times form the
sequence $\{1,2,3,4,5,6,7,8\}$ of length $N=8$. Now let us read the deposition times of the
dots sequentially, but now column by column and vertically from top to bottom
in each column, starting from the leftmost one. Then this sequence reads
$\{8,3,5,1,2,6,4,7\}$ which is just a permutation of the original sequence
$\{1,2,3,4,5,6,7,8\}$. In the permuted sequence $\{8,3,5,1,2,6,4,7\}$ there are
$3$ LIS's: $\{3,5,6,7\}$, $\{1,2,6,7\}$ and $\{1,2,4,7\}$, all of the same
length $l_N=4$. As mentioned before (see Fig. \ref{psheap}), this is precisely
the number of piles in the patience sorting of the permutation
$\{8,3,5,1,2,6,4,7\}$.

Let us note one immediate fact from Fig. \ref{fig:3}. The numbers
belonging to the different level lines in Fig. \ref{fig:3} are in one-to-one
correspondence with the piles $[\{8,3,1\}, \{5,2\}, \{6,4\},\{7\}]$ in
Aldous--Diaconis patience sorting game. Hence, each pile can be identified with
an unique level line. Now, the height $h(x,t)$ at any given point $(x,t)$ in
Fig. \ref{fig:3} is equal to the number of level lines inside the rectangle
bounded by the corners: $[0,0], [x,0], [0,t], [x,t]$. Thus, we have
the correspondonce: height $\equiv$ number of level lines $\equiv$ number of piles $\equiv$
length $l_n$ of the LIS. However, to compute $l_n$, we need to know the value of $n$ which
is precisely the number of black dots inside this rectangle.

Once the problem is reduced to finding the number of black dots or deposition events, we
no longer need the Fig. \ref{fig:3} (as it may confuse due to the visual shift
$(x,y)\to (x+h(x,y),y)$) and can go back to Fig. \ref{fig:2}, where the
north-to-east corners play the same role as the black dots in Fig. \ref{fig:2}.
In Fig. \ref{fig:2}, to determine the height $h_k(t)$ of the $k$-th column at
time $t$, we need to know the number of deposition events inside the $2$D plane
rectangle $R_{k,t}$ bounded by the four corners $[0,0], [k,0], [0,t], [k,t]$.
Let us begin with the last column $k=L$. For $k=L$ the number of deposition
events $N$ in the rectangle $R_{L,t}$ is equal to the time $t$ because there is
only one deposition event per time. In our example $N=t=8$. For a general $k<L$
the number of deposition events $N$ inside the rectangle $R_{k,t}$ is a random
variable, since some of the rows inside the rectangle may not contain a
north-to-east corner or a deposition event. The probability distribution
$P_{k,t}(N)$ (for a given $[k,t]$) of this random variable can, however, be
easily found as follows. At each step of deposition, a column is chosen at
random from any of the $L$ columns. Thus, the probability that a north-to-east
corner will fall on the segment of line $[0,k]$ (where $k\leq L$) is equal to
$k/L$. The deposition events are completely independent of each other,
indicating the absence of correlations between different rows labelled by $t$ in
Fig. \ref{fig:2}. So, we are asking the question: given $t$ rows, what is the
probability that $N$ of them will contain a north-to-east corner? This is
simply given by the binomial distribution
\begin{equation}
P_{k,t}(N) = {t\choose N } \left({\frac {k}{L}} \right)^N
\left(1-{\frac {k}{L}}\right)^{t-N},
\label{binom1}
\end{equation}
where $N\leq t$. Now we are reduced to the following problem: given a sequence
of integers of length $N$ (where $N$ itself is random and is taken from the
distribution in Eq.(\ref{binom1})), what is the length of the LIS? Recall that
this length is precisely the height $h_k(t)$ of the $k$-th column at time $t$
in our model. In the thermodynamic limit $L\to \infty$ for $t\gg 1$ and any
fixed $k$ such that the quotient $\lambda=\frac{tk}{L}$ remains fixed but is
arbitrary, the distribution in Eq.(\ref{binom1}) becomes a Poisson distribution
$P(N)\to e^{-\lambda} \frac {\lambda^N}{N!}$, with the mean
$\lambda=\frac{tk}{L}$. We can then directly use the BDJ result in
Eq.(\ref{bdj1}) to predict our main result for the height in the BD model,
\begin{equation}
h_k(t) \to 2\sqrt{\frac{tk}{L}} + \left(\frac{tk}{L}\right)^{1/6} \chi,
\label{result1}
\end{equation}
for large $\lambda=tk/L$, where the random variable $\chi$ has the
Tracy-Widom distribution $F_2(\chi)$ as in Eq. (\ref{gue}). 
Using the known exact value $\langle \chi\rangle
=-1.7711...$ from the Tracy-Widom distribution \cite{TW1}, we find exactly the
asymptotic average height profile in the BD model,
\begin{equation}
\langle h_k(t)\rangle \to 2\sqrt{\frac{tk}{L}}-
1.7711...\left(\frac{tk}{L}\right)^{1/6}.
\label{avgh}
\end{equation}
The leading square root dependence of the profile on the column number $k$ has
been seen numerically. Eq. (\ref{avgh}) also predicts an
exact sub-leading term with $k^{1/6}$ dependence. Similarly, for the variance,
$\sigma_k^2(t)=\langle [h_k(t)-\langle h_k(t)\rangle]^2 \rangle$, we find
asymptotically: $\sigma_k^2(t)\to c_0\left(\frac{tk}{L}\right)^{1/3}$, where
$c_0=\langle [\chi-\langle \chi \rangle]^2\rangle=0.8132...$ \cite{TW1}.
Eliminating the $t$ dependence for large $t$ between the average and the
variance, we get, $\sigma_k^2(t)\approx a {\langle h_k(t)\rangle}^{2\beta}$
where the constant $a=c_0/2^{2/3}=0.51228\dots$ and $\beta=1/3$, thus
recovering the KPZ scaling exponent.
In addition to the BD model with infinite range right-handed
interaction reported here,
we have also analyzed the model (analytically within a mean field theory and numerically)
when the right-handed interaction is short ranged.
Somewhat suurprisingly and pleasantly, we found that
the asymptotic average height profile is independent of the range of interaction.
A recent analysis of the short range BD model sheds light on this fact~\cite{KNV}.

So far, we have demonstrated that for a flat initial condition, the height fluctuations in the
BD model follow the Tracy-Widom distribution $F_{\rm GUE}(x)$ which corresponds to
the distribution of the largest eigenvalue of a random matrix drawn from a Gaussian unitary ensemble.
In the context of the PNG model, Pr\"ahofer and Spohn \cite{PS} have shown that while the height
fluctuations of a single PNG droplet follow the distribution $F_{\rm GUE}(x)$, it is possible to
obtain other types of universal distributions as well. For example, the height fluctuations
in the PNG model growing over a flat substrate follow the
distribution $F_{\rm GOE}(x)$ where $F_{\rm GOE}(x)$ is the distribution of the largest
eigenvalue of a random matrix drawn from the Gaussian orthogonal ensemble. Besides,
in a PNG droplet with two external sources at its edges which nucleate with rates
$\rho_{+}$ and $\rho_{-}$, the height fluctuations have different distributions depending
on the values of $\rho_{+}$ and $\rho_{-}$. For $\rho_{+}<1$ and $\rho_{-}<1$, one gets back
the distribution $F_{\rm GUE}(x)$. If however $\rho_{+}=1$ and $\rho_{-}<1$ (or alternatively
$\rho_{-}=1$ and $\rho_{+}<1$), one gets the distribution $F_{\rm GOE}^2(x)$ which corresponds to
the distribution of the largest of the superimposed eigenvalues of two independent
GOE matrices. In the critical case $\rho_{+}=1$ and $\rho_{-}=1$, one gets a new
distribution $F_0(x)$ which does not have any random matrix analogy. For $\rho_{+}>1$
and $\rho_{-}>1$, one gets Gaussian distribution. These results for the PNG model were obtained in
Ref. \cite{PS} using a powerful theorem of Baik and Rains \cite{BR1}.

The question naturally arises as to whether these other distributions, apart from the $F_{\rm GUE}(x)$,
can also appear in the BD model considered in this paper. Indeed, they do. For example, if
one starts with a staircase initial condition $h_k(0)=k$ for the heights in the BD model,
one gets the distribution $F_{\rm GOE}^2(x)$ for the scaled variable $\chi$. This follows from the
fact that for the staircase initial condition, in Fig. 2 there will be a black dot (or a north-to-east
corner) at every value of $k$ on the $k$ axis at $t=0$. Thus the black dots appear on the $k$ axis
with unit density. This
corresponds to the case $\rho_{+}=1$
and $\rho_{-}=0$ of the general results of Baik and Rains which leads to a $F_{\rm GOE}^2(x)$
distribution. Of course, the density $\rho_{+}$ can be tuned between $0$ and $1$, by tuning
the average slope of the staircase. For a generic $0<\rho_{+}\leq 1$, one can also
vary $\rho_{-}$ by putting an external source at the first column.
Thus one can obtain, in principle, most of the distributions discussed in Ref. \cite{BR1} by varying
$\rho_{+}$ and $\rho_{-}$.
Note that the
case $\rho_{-}=1$ (external source which drops one particle at the first column at every time step) and
$\rho_{+}=0$ (flat substrate) is, however, trivial since the surface then remains flat
at all times and the height just increases by one unit at every time step. The distribution
$F_{\rm GOE}(x)$ is, however, not naturally accessible within the rules of our model.

\section{Sequence Matching Problem}

In this section, I will discuss a different problem namely that of the alignment of two
random sequences and will illustrate how the Tracy-Widom distribution appears in this 
problem. This is based on a joint wotk with S. Nechaev~\cite{MN}.

\begin{figure}
\includegraphics[width=.7\hsize]{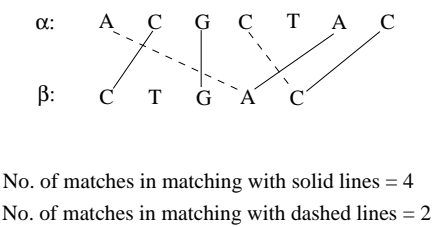}
\caption{ Two fixed sequences $\alpha: \{A, C, G, C, T, A, C\}$
and $\beta: \{C, T, G, A, C\}$. The solid lines show the common
subsequence $\{C, G, A, C\}$ and the dashed lines denote another
common subsequence $\{A,C\}$.}
\label{matching}
\end{figure}

Sequence alignment is one of the most useful quantitative methods used in
evolutionary molecular biology\cite{W1,Gusfield,DEKM}. The goal of an alignment
algorithm is to search for similarities in patterns in different sequences. A
classic and much studied alignment problem is the so called `longest common
subsequence' (LCS) problem. The input to this problem is a pair of sequences
$\alpha=\{\alpha_1, \alpha_2,\dots, \alpha_i\}$ (of length $i$) and
$\beta=\{\beta_1, \beta_2,\dots, \beta_j\}$ (of length $j$). For example, $\alpha$
and $\beta$ can be two random sequences of the $4$ base pairs $A$, $C$, $G$, $T$ of
a DNA molecule, e.g., $\alpha=\{A, C, G, C, T, A, C\}$ and $\beta=\{C, T, G, A,
C\}$. A subsequence of $\alpha$ is an ordered sublist of $\alpha$ (entries of which
need not be consecutive in $\alpha$), e.g, $\{C, G, T, C\}$, but not $\{T, G, C\}$.
A common subsequence of two sequences $\alpha$ and $\beta$ is a subsequence of both
of them. For example, the subsequence $\{C, G, A, C\}$ is a common subsequence of
both $\alpha$ and $\beta$. There can be many possible common subsequences of a pair
of sequences. For example, another common subsequence of $\alpha$ and $\beta$ is
$\{A, C\}$. One simple way to construct different common subsequences (for two
fixed sequences $\alpha$ and $\beta$) is by drawing lines from one member
of the set $\alpha$ to another member of the set $\beta$ such that the lines
can not cross. For example, the common subsequence $\{C, G, A, C\}$ is shown
by solid lines in Fig. \ref{matching}. On the other hand the common subsequence
$\{A,C\}$ is shown by the dashed lines in Fig. \ref{matching}.  
The aim of the LCS problem is to find the longest of such common
subsequences between two fixed sequences $\alpha$ and $\beta$. 

This problem and its variants have been widely studied in
biology\cite{NW,SW,WGA,AGMML}, computer science\cite{SK,AG,WF,Gusfield}, probability
theory\cite{CS,Deken,Steele,DP,Alex,KLM} and more recently in statistical
physics\cite{ZM,Hwa,Monvel}. A particularly important application of the LCS problem
is to quantify the closeness between two DNA sequences. In evolutionary biology, the
genes responsible for building specific proteins evolve with time and by finding the
LCS of the same gene in different species, one can learn what has been conserved in
time. Also, when a new DNA molecule is sequenced {\it in vitro}, it is important to
know whether it is really new or it already exists. This is achieved quantitatively
by measuring the LCS of the new molecule with another existing already in the
database.

For a pair of fixed sequences of length $i$ and $j$ respectively, the length
$L_{i,j}$ of their LCS is just a number. However, in the stochastic version of the
LCS problem one compares two random sequences drawn from $c$ alphabets and hence the
length $L_{i,j}$ is a random variable. A major challenge over the last three decades
has been to determine the statistics of $L_{i,j}$\cite{CS,Deken,Steele,DP,Alex}. For
equally long sequences ($i=j=n$), it has been proved that $\langle L_{n,n}\rangle
\approx \gamma_c n$ for $n\gg 1$, where the averaging is performed over all
realizations of the random sequences. The constant $\gamma_c$ is known as the
Chv\'atal-Sankoff constant which, to date, remains undetermined though there exists
several bounds\cite{Deken,DP,Alex}, a conjecture due to Steele\cite{Steele} that
$\gamma_c=2/(1+\sqrt{c})$ and a recent proof\cite{KLM} that $\gamma_c\to 2/\sqrt{c}$
as $c\to \infty$. Unfortunately, no exact results are available for the finite size
corrections to the leading behavior of the average $\langle L_{n,n}\rangle$, for the
variance, and also for the full probability distribution of $L_{n,n}$. Thus, despite
tremendous analytical and numerical efforts, exact solution of the random LCS
problem has, so far, remained elusive. Therefore it is important to find other
variants of this LCS problem that may be analytically tractable.

Computationally, the easiest way to determine the length $L_{i,j}$ of the LCS of two
arbitrary sequences of lengths $i$ and $j$ (in polynomial time $\sim O(ij)$) is via
using the recursive algorithm\cite{Gusfield,Monvel}
\begin{equation}
L_{ij} = \max\left[L_{i-1,j}, L_{i,j-1}, L_{i-1,j-1} + \eta_{i,j}\right],
\label{recur1}
\end{equation}
subject to the initial conditions $L_{i,0}=L_{0,j}=L_{0,0}=0$. The variable
$\eta_{i,j}$ is either 1 when the characters at the positions $i$ (in the sequence
$\alpha$) and $j$ (in the sequence $\beta$) match each other, or 0 if they do not.
Note that the variables $\eta_{i,j}$'s are not independent of each other. To see
this consider the simple example -- matching of two strings $\alpha={\rm AB}$ and
$\beta={\rm AA}$. One has by definition: $\eta_{1,1}=\eta_{1,2}=1$ and
$\eta_{2,1}=0$. The knowledge of these three variables is sufficient to predict that
the last two letters will not match, i.e., $\eta_{2,2}=0$. Thus, $\eta_{2,2}$ can
not take its value independently of $\eta_{1,1},\,\eta_{1,2},\,\eta_{2,1}$. These
residual correlations between the $\eta_{i,j}$ variables make the LCS problem rather
complicated. Note however that for two random sequences drawn from $c$ alphabets,
these correlations between the $\eta_{i,j}$ variables vanish in the $c\to \infty$
limit.

A natural question is how important are these correlations between the $\eta_{i,j}$ variables, e.g.,
do they affect the asymptotic statistics of $L_{i,j}$'s for large $i$ and $j$?
Is the problem solvable if one ignores these correlations?
These questions naturally lead to the Bernoulli matching (BM) model which is a simpler variant of
the original LCS problem where one ignores the correlations between $\eta_{i,j}$'s for all
$c$\cite{Monvel}.
The length $L_{i,j}^{BM}$ of the BM model satisfies the same
recursion relation in Eq. (\ref{recur1}) except that $\eta_{i,j}$'s are now
independent and each drawn from the bimodal distribution: $p(\eta)=
(1/c)\delta_{\eta,1}+ (1-1/c)\delta_{\eta,0}$.
This approximation is expected to be exact only in the appropriately taken
$c\to \infty$ limit. Nevertheless, for finite $c$, the results on the BM model can serve
as a useful benchmark for the original LCS model to decide if indeed the correlations
between $\eta_{i,j}$'s are important or not. Unfortunately, even in the absence of
correlations, the exact aymptotic distribution of $L_{i,j}^{BM}$ in the BM model has so far
remained elusive, mainly due to the nonlinear nature of the recursion relation
in Eq. (\ref{recur1}).
The purpose of this Rapid Communication is to present an exact asymptotic formula for the
distribution of the length $L_{n,n}^{BM}$ in the BM model for all $c$.
So far, only the leading asymptotic behavior of the
average length in the BM model is known\cite{Monvel} using the `cavity'
method of spin glass physics\cite{MPV},
\begin{equation}
\langle L_{n,n}^{BM}\rangle  \approx \gamma_c^{BM} n
\label{bm1}
\end{equation}
where $\gamma_c^{BM}= 2/(1+\sqrt{c})$, same as the conjectured value of the
Chv\'atal-Sankoff constant $\gamma_c$ for the original LCS model. However, other
properties such as the variance or the distribution of $L_{n,n}^{BM}$ remained
untractable even in the BM model.
We have shown~\cite{MN}, as illustrated below, that for large $n$,
\begin{equation}
L_{n,n}^{BM}\to \gamma_c^{BM} n + f(c)\, n^{1/3}\, \chi 
\label{asymp11}
\end{equation}
where $\chi$ is a random variable with a $n$-independent distribution, ${\rm Prob}
(\chi\le x)= F_{ 2}(x)$ which is precisely the Tracy-Widom distribution
in Eq. (\ref{gue}). 
Indeed, we were also able to compute the functional form of the scale factor $f(c)$ exactly for all 
$c$~\cite{MN},
\begin{equation}
f(c)=\frac{c^{1/6}(\sqrt{c}-1)^{1/3}}{\sqrt{c}+1}.
\label{fc1}
\end{equation}
This allows us to calculate the average including the subleading finite size
correction term and the variance of $L_{n,n}^{BM}$ for large $n$,
\begin{eqnarray}
\langle L_{n,n}^{BM}\rangle &\approx & \gamma_c^{BM} n + \left<\chi\right> f(c)
n^{1/3} \nonumber \\
{\rm Var}\, L_{n,n}^{BM} &\approx &
\left(\langle\chi^2\rangle-{\langle\chi\rangle}^2\right)\, f^2(c)\, n^{2/3},
\label{eq:expvar}
\end{eqnarray}
where one can use the known exact values\cite{TW1}, $\langle \chi\rangle=
-1.7711\dots$ and $\langle \chi^2\rangle- {\langle \chi\rangle}^2= 0.8132\dots$.
These exact results thus invalidate the previous attempt\cite{Monvel} to
fit the subleading correction to the mean in the BM model with a
$n^{1/2}/{\ln (n)}$ behavior and also to fit the scaled distribution
with a Gaussian form.
Note that the recursion relation in Eq.
(\ref{recur1}) can also be viewed as a $(1+1)$ dimensional directed polymer
problem\cite{Hwa,Monvel} and some asymptotic results (such as the $O(n^{2/3})$
behavior of the variance of $L_{n,n}$ for large $n$) can be obtained using the
arguments of universality\cite{Hwa}. However, this does not provide precise results
for the full distribution along with the correct scale factors that are obtained here.

It is useful to provide a synopsis of our method in deriving these results. First,
we prove the results in the $c\to \infty$ limit, by using mappings to other models.
To make progress for finite $c$, we first map the BM model exactly to a $3$-d
anisotropic directed percolation (ADP) model first studied by Rajesh and
Dhar\cite{RD}. This ADP model is also precisely the same as the directed
polymer model studied by Johansson~\cite{J1}, as discussed in the previous section
and for which the exact results are known as in Eq. (\ref{j1}).
To extract the results for the BM model from those of Johansson's
model, we use a simple symmetry argument which then allows us to derive our main
results in Eqs. (\ref{asymp11})-(\ref{eq:expvar}) for all $c$. As a check, we recover
the $c\to \infty$ limit result obtained independently by the first method.

\begin{figure}
\includegraphics[width=.7\hsize]{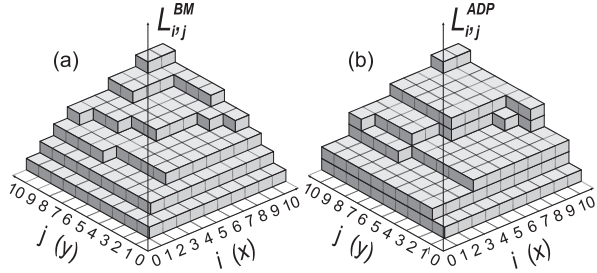}
\caption{Examples of (a) BM surface
$L_{i,j}^{BM}\equiv {\tilde h}(x,y)$ and (b) ADP surface $L_{i,j}^{ADP}\equiv
h(x,y)$.} 
\label{fig:bms1}
\end{figure}

In the BM model, the length $L_{i,j}^{BM}$ can be interpreted as the height of a
surface over the $2$ dimensional $(i,j)$ plane constructed via the recursion relation in Eq.
(\ref{recur1}). A typical surface, shown in Fig. \ref{fig:bms1}\,(a), has terrace-like structures.

\begin{figure}
\includegraphics[width=.7\hsize]{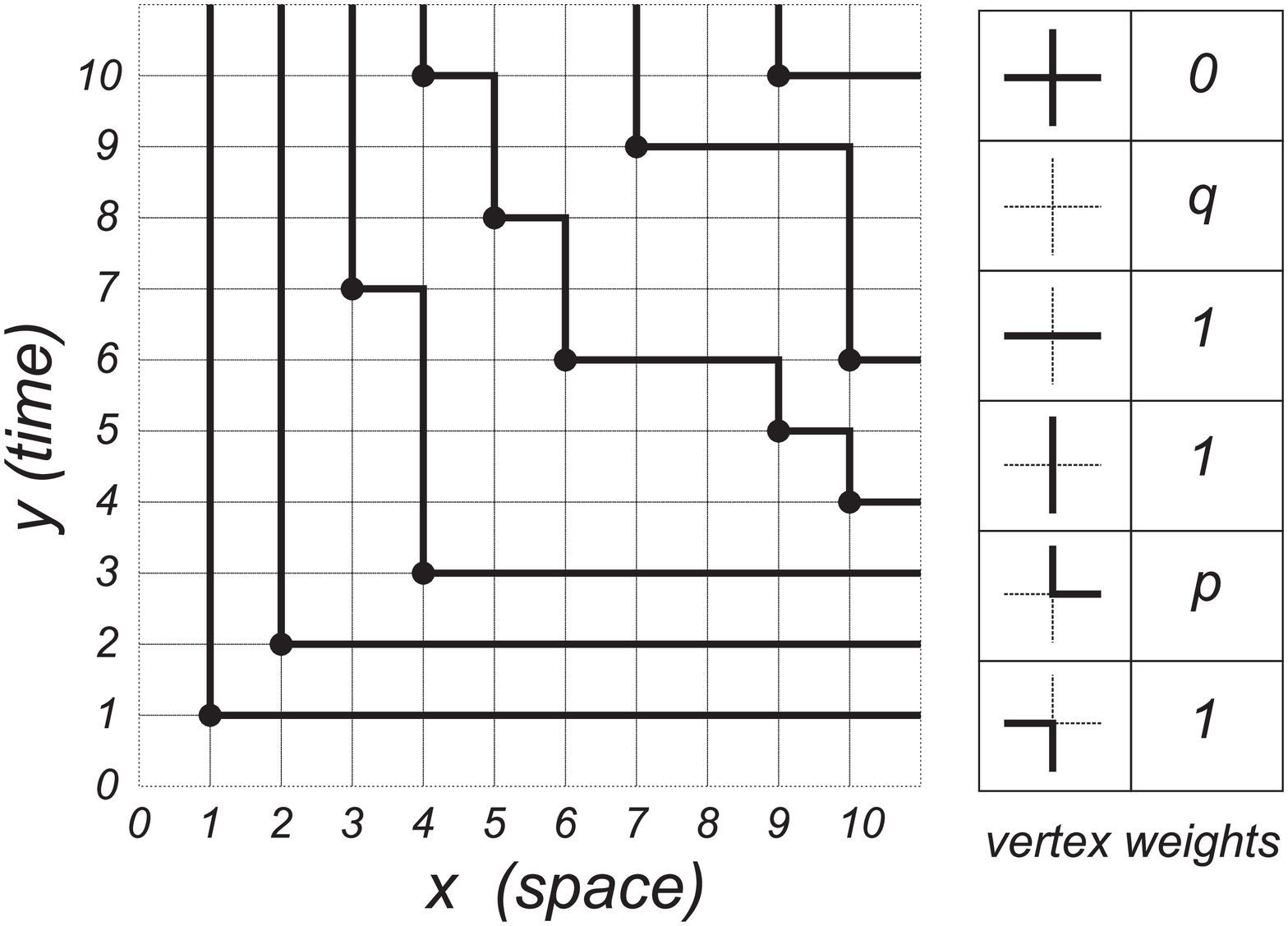}
\caption{Projected $2$-d level lines separating adjacent terraces of unit height
difference in the BM surface in Fig.\ref{fig:bms1} (a). The adjacent table shows the weights of
all vertices on the $2$-d plane.} 
\label{fig:bms2}
\end{figure}

It is useful to consider the projection of the level lines separating the adjacent
terraces whose heights differ by $1$ (see Fig.\ref{fig:bms2}) onto the $2$-D $(i,j)$ plane. Note
that, by the rule Eq. (\ref{recur1}), these level lines never overlap each other,
i.e., no two paths have any common edge. The statistical weight of such a projected
$2$-D configuration is the product of weights associated with the vertices of the
$2$-D plane. There are five types of possible vertices with nonzero weights as shown
in Fig. \ref{fig:bms2}, where $p=1/c$ and $q=1-p$. Since the level lines never cross each other,
the weight of the first vertex in Fig. \ref{fig:bms2} is $0$.

Consider first the limit $c\to \infty$ (i.e., $p\to 0$). The weights of all allowed
vertices are $1$, except the ones shown by black dots in Fig. \ref{fig:bms2}, whose associated
weights are $p\to 0$. The number $N$ of these black dots inside a rectangle of area
$A=ij$ can be easily estimated.
For large $A$ and $p\to 0$, this number
is clearly
Poisson
distributed with the mean ${\overline N}= pA$.
The height $L_{i,j}^{BM}$ is just the number of level lines $\cal N$ inside this
rectangle of area $A=ij$. One can easily estimate $\cal N$ by following
precisely the method outlined in the previous subsection in the context of the ballistic deposition
model. Following the same analysis as in the ballistic deposition model,
it is easy to see that
the number of level lines ${\cal N}$ inside the rectangle
(for large $A$), appropriately scaled, has a limiting behavior, ${\cal N}\to
2\sqrt{\overline N} + {\overline N}^{1/6}\, \chi$, where $\chi$ is a random variable
with the Tracy-Widom distribution. Using ${\overline N}=pA=ij/c$, one then obtains in
the limit $p\to 0$,
\begin{equation}
L_{i,j}^{BM}= {\cal N} \to \frac{2}{\sqrt c}\sqrt{ij} +
{\left( \frac{ij}{c}\right)}^{1/6}\, \chi.
\label{p01}
\end{equation}
In particular, for large equal length sequences $i=j=n$, we get for $c\to \infty$
\begin{equation}
L_{n,n}^{BM}\to \frac{2}{\sqrt{c}}\, n + c^{-1/6} \, n^{1/3}\, \chi .
\label{p02}
\end{equation}
For finite $c$, while the above mapping to the LIS problem still works, the
corresponding permutations of the LIS problem are not generated with equal
probability and hence one can no longer use the BDJ results.

For any finite $c$, we can however map the BM model to the ADP model studied by Rajesh and Dhar~\cite{RD}. 
In the ADP model on
a simple cubic lattice the bonds are occupied with probabilities $p_x$, $p_y$, and
$p_z$ along the $x$, $y$ and $z$ axes and are all directed towards increasing
coordinates. Imagine a source of fluid at the origin which spreads along the
occupied directed bonds. The sites that get wet by the fluid form a $3$-d cluster.
In the ADP problem, the bond occupation probabilities are anisotropic, $p_x=p_y=1$
(all bonds aligned along the $x$ and $y$ axes are occupied) and $p_z=p$. Hence, if
the point $(x,y,z)$ gets wet by the fluid then all the points $(x',y', z)$ on the
same plane with $x'\ge x$ and $y'\ge y$ also get wet. Such a wet cluster is compact
and can be characterized by its bounding surface height $H(x,y)$ as shown in
Fig.(1b). It is not difficult to see~\cite{RD} that the height $H(x,y)$ satisfies exactly
the same recursion relation of the directed polymer as in Eq. (\ref{dpr2})
where $\xi_{x,y}$'s are i.i.d. random variables taking nonnegative integer values
with ${\rm Prob}(\xi_{x,y}=k)= (1-p)\, p^k$ for $k=0,1,2,\dots$. Thus the ADP
model of Rajesh and Dhar is precisely identical to the directed polymer model
studied by Johansson with exactly the same distribution of the noise $\xi(x,y)$.

While the terrace-like structures of the ADP surface look similar to the BM surfaces
(compare Figs. (\ref{fig:bms1}\,a) and (\ref{fig:bms1}\,b), there is an important difference between the 
two. In
the ADP model, the level lines separating two adjacent terraces can overlap with
each other\cite{RD}, which does not happen in the BM model. However, by making the
following change of coordinates in the ADP model\cite{RD}
\begin{equation}
\zeta= x+ h(x,y); \,\,\, \eta=y+ h(x,y)
\label{ct1}
\end{equation}
one gets a configuration of the surface where the level lines no longer overlap.
Moreover, it is not difficult to show that the projected $2$-D configuration of
level lines of this shifted ADP surface has exactly the same statistical weight as
the projected $2$-D configuration of the BM surface. Denoting the BM height by
${\tilde h}(x,y)= L_{x,y}^{BM}$, one then has the identity, ${\tilde h}(\zeta,
\eta)= h(x,y)$, which holds for each configuration. Using Eq. (\ref{ct1}), one can
rewrite this identity as
\begin{equation}
{\tilde h}(\zeta, \eta)= h\left( \zeta- {\tilde h}(\zeta, \eta),
\eta- {\tilde h}(\zeta, \eta)\right).
\label{conv1}
\end{equation}

Thus, for any given height function $h(x,y)$ of the ADP model, one can, in
principle, obtain the corresponding height function ${\tilde h}(x,y)$ for all
$(x,y)$ of the BM model by solving the nonlinear equation (\ref{conv1}). This is
however very difficult in practice. Fortunately, one can make progress for large
$(x,y)$ where one can replace the integer valued discrete heights by continuous
functions $h(x,y)$ and ${\tilde h}(x,y)$. Using the notation $\partial_x\equiv
\partial/{\partial x}$ it is easy to derive from Eq. (\ref{ct1}) the following pair
of identities,
\begin{equation}
\partial_x h = \frac{\partial_{\zeta} {\tilde h}}{1-\partial_{\zeta}
{\tilde h}-\partial_{\eta} {\tilde h}};
\,\,\,
\partial_y h = \frac{\partial_{\eta} {\tilde h}}{1-\partial_{\zeta}
{\tilde h}-\partial_{\eta} {\tilde h}}.
\label{der1}
\end{equation}
In a similar way, one can show that
\begin{equation}
\partial_{\zeta} {\tilde h} = \frac{\partial_x h}{1+\partial_x h+\partial_y h};\,\,\,
\partial_{\eta} {\tilde h} = \frac{\partial_y h}{1+\partial_x h+\partial_y h}.
\label{der2}
\end{equation}
We then observe that Eqs. (\ref{der1}) and (\ref{der2}) are invariant under the
simultaneous transformations
\begin{equation}
\zeta\to -x ; \,\, \eta\to -y; \,\, \tilde h \to h \, .
\label{invar1}
\end{equation}
Since the height is built up by integrating the derivatives, this leads to a simple
result for large $\zeta$ and $\eta$,
\begin{equation}
{\tilde h}(\zeta, \eta) = h(-\zeta, -\eta).
\label{res1}
\end{equation}

Thus, if we know exactly the functional form of the ADP surface $h(x,y)$, then the
functional form of the BM surface ${\tilde h}(x,y)$ for large $x$ and $y$ is simply
obtained by ${\tilde h}(x,y)=h(-x,-y)$. Changing $x\to -x$ and $y\to -y$ in
Johansson's expression for the ADP surface in Eq. (\ref{j1}) we thus arrive at our
main asymptotic result for the BM model
\begin{eqnarray}
L_{x,y}^{BM}&=& {\tilde h}(x,y) \to \frac{2\sqrt{pxy}-p(x+y)}{q}+ \nonumber \\
&+&\frac{(pxy)^{1/6}}{q}\,\left[(1+p)-\sqrt{\frac{p}{xy}}\,(x+y)\right]^{2/3} \,
\chi, \label{res2}
\end{eqnarray}
where $p=1/c$ and $q=1-1/c$. For equal length sequences $x=y=n$, Eq. (\ref{res2})
then reduces to Eq. (\ref{asymp11}).

To check the consistency of our asymptotic results, we further computed the
difference between the left- and the right-hand sides of Eq. (\ref{conv1}),
\begin{equation}
\Delta h (\zeta, \eta)= {\tilde h}(\zeta, \eta)- h\left( \zeta- {\tilde h}(\zeta,
\eta), \eta- {\tilde h}(\zeta, \eta)\right), \label{conv2}
\end{equation}
with the functions $h(x,y)$ and ${\tilde h}(x,y)$ given respectively by Eqs.
(\ref{j1}) and (\ref{res2}). For large $\zeta=\eta$ one gets
\begin{equation}
\Delta h(\zeta,\zeta) \to \left[{p^{1/3}\chi^2}/{3 (1-\sqrt{p})^{4/3}}\right]\,
{\zeta}^{-1/3} . \label{cons1}
\end{equation}
Thus the discrepancy falls off as a power law for large $\zeta$, indicating that
indeed our solution is asymptotically exact. We have also performed numerical
simulations of the BM model using the recursion relation in Eq. (\ref{recur1}) for
$c=2,\,4,\,9,\,16,\,100$. Our preliminary results\cite{MN} for relatively small
system sizes (up to $n=5000$) are consistent with our exact results in Eqs.
(\ref{asymp11})-(\ref{eq:expvar}).

Thus, the Tracy-Widom distribution also describes the asymptotic distribution of
the optimal matching length in the BM model, for all $c$. Given that the correlations in the original LCS 
model
become negligible in the $c\to \infty$ limit, it is likely that the
BM asymptotics in Eq. (\ref{p02}) would also hold for the original LCS model
in the $c\to \infty$ limit.
An important open problem
is to determine whether the Tracy-Widom distribution also appears in the
LCS problem for finite $c$. The precise distribution obtained
here (including exact prefactors) for all $c$ in the BM model will serve
as a useful benchmark to which future simulations of the LCS problem can
be compared.

\section{Conclusion}

In these lectures I have discussed $4$ a priori unrelated problems and tried to give a flavour
of the recent developments that have found a deep connection between these problems.
These connections have now established the fact that they all share one common limiting distribution,
namely the Tracy-Widom distribution that describes the asymptotic distribution law of
the largest eigenvalue of a random matrix. I have also discussed the probabilities of
large deviations of the largest eigenvalue, in the range outside the validity of the 
Tracy-Widom law. As examples, I have demonstrated in detail, in two specfic models
a ballistic
deposition model and a sequence alignment problem, 
how they can be mapped on to the longest increasing subsequence problem
and consequently proving the existence of the Tracy-Widom distribution in these
models.

There have been many other interesting recent developments in this rather broad area encompassing
different fields that I did not have the scope to discuss in these lectures.
There are, of course, plenty of open questions that
need to be addressed, some of which I mention below.

{\em Finite size effects in growth models:} We have discussed how the Tracy-Widom distribution appears 
as the limiting scaled height distribution in several $(1+1)$ dimensional growth
models that belong to the KPZ universality class of fluctuating interfaces. Indeed,
for a fluctuating surface with height $H(x,t)$ growing over a substrate of infinite size 
one now believes that at long times $t>>1$
\begin{equation}
H(x,t) = v t + b t^{1/3} \chi 
\label{con1}
\end{equation}
where $\chi$ is a time-independent random variable with the Tracy-Widom distribution.
The prefactors $v$ (the velocity of the interface) and $b$ are model dependent,
but the distribution of the scaled variable $\chi=(H-vt)/{bt^{1/3}}$ is universal
for large $t$. The nonuniversal prefactors are often very hard to compute. We have
shown two examples in these lectures where these prefactors can be computed exactly. 
Note, however, that the result in Eq. (\ref{con1}) holds only in an infinite system. 
In any real system with a finite
substrate size $L$, the result in Eq. (\ref{con1}) will hold only in the growing
regime of the surface, i.e., when $1<< t << L^z$, where $z$ is the dynamical
exponent characterizing the surface evolution. For example, for the KPZ
type of interfaces in $(1+1)$ dimensions, $z=3/2$. However, when $t>> L^z$, the probability distribution
of the height fluctuation 
$H-\langle H\rangle$ will become time-independent. For example, for $(1+1)$ dimensional KPZ surfaces
with periodic boundary conditions, it is well known~\cite{HZ} that the stationary distribution of
the height fluctuation is a simple Gaussian, ${\rm Prob}[H-\langle H\rangle=x]\propto \exp[-x^2/{a_0 L}]$
where $a_0$ is a nonuniversal constant and the typical fluctuation scales with the system size as $L^{1/2}$.
An important open question is how does the distribution of the height fluctuation crosses over
from the Tracy-Widom form to a simple Gaussian form as $t$ becomes bigger than the crossover time $L^z$.
It would be nice to show this explicitly in any of the simple models discussed above.

{\em A direct connection between the growth models and random matrices:} The existence of the Tracy-Widom
distribution in many of the growth models discussed here, such as the polynuclear growth model
or the ballistic deposition model, rely on the mapping to the LIS problem
and subsequently using the BDJ results that connect the LIS problem to random matrices.
It is certainly desirable to find to a direct mapping between the growth models and the
largest eigenvalue of a random matrix. Recent work by Spohn and collaborators~\cite{Spohn}
linking the top edge of a PNG growth model to Dyson's brownian motion of the eigenvalues
of a random matrix perhaps provides a clue to this missing link.

{\em Largest Lyapunov exponent in population dynamics:} The Tracy-Widom distribution
and the associated large-deviation function discussed in Section 3 
conceivably have important applications in several systems
where the largest eigenvalue controls the spectral properties of the system. Some
examples were discussed in Section 3. Recently, it has been shown that the statistics
of largest eigenvalue (the largest Lyapunov exponent) is also of importance
in population growth of organisms in fluctuating environments~\cite{KL1}.  
It would be interesting to see if Tracy-Widom type distribution functions also 
appear in these biological problems.

{\em Sequence matching, directed polymer and vertex models:} In the context of the sequence matching problem
discussed in Section 6, we have demonstrated how the statistical weights of the surface generated
in the Bernoulli matching 
model of the sequence alignment are exactly identical to that of 
a $5$-vertex model on a square lattice (see Fig. \ref{fig:bms2}). This is a useful connection
because there are many quantities in the $5$-vertex models that can be computed exactly by employing
the Bethe ansatz techniques and subsequently one can use those results for the sequence
alignment or equivalently for the directed polymer model. Recently, in collaboration
with K. Mallick and S. Nechaev, we have made some progress in these directions~\cite{MMN}.
A very interesting open issue is if one can derive the Tracy-Widom distribution 
by using the Bethe ansatz techniques.
    
{\em Other issues related to the sequence matching problem:} There are also many other
interesting open questions associated with 
the sequence matching problem.
We have shown that the length of the longest matching is Tracy-Widom distributed
only in the Bernoulli matching model which is a simpler version of the original LCS problem.
In the BM model one has ignored certain correlations, as we discussed in detail. This approximation is
exact in the $c\to \infty$ limit, where $c$ is the number of different types of alphabets, e.g.
for DNA, $c=4$. Is this approximation good even for finite $c$? In 
other words,
is the optimal matching length in the original LCS problem also Tracy-Widom distributed?
It would also be
interesting if one can make a systematic $1/c$ expansion of the LCS model, i.e., keeping
the correlations up to $O(1/c)$. Numerical simulations the LCS problem~\cite{BMat} for binary sequence $c=2$
indeed indicates that the standard deiviation of the optimal matching length scales as $n^{1/3}$ where
$n$ is the sequence size, as in the
BM model, the question is if the scaled distribution is also Tracy-Widom or not. 
For the original LCS problem, there is also a curious result due to Bonetto
and Matzinger~\cite{BMat} that claims that if the value of $c$ for the two sequences are not the same (for example,
the first sequence may be drawn randomly from $3$ alphabets and the second may be a binary sequence),
then the standard deviation of the optimal matching length scales as $n^{1/2}$ for large $n$, which
is rather surprising!
It would be interesting to study the statistics of optimal matches between more than two sequences.
Finally, here we have just mentioned the matching of random sequences. It would be interesting
and important
to study the statistics of optimal matching lengths between non-random sequences, e.g.,
when there are some correlations between the members of any given sequence.

\vspace{0.2cm}

{\bf Acknowledgements:} My own contribution to this field that is presented here was 
developed partly in collaboration
with D.S. Dean and partly with S. Nechaev. It is a pleasure to thank them.
I also thank O. Bohigas, K. Mallick and P. Vivo for collaborations on related topics.
Besides, I acknowledge useful discussions with G. Biroli, J.-P. Bouchaud, A.J. Bray, 
A. Comtet, D. Dhar, S. Leibler, O.C. Martin, M. M\'ezard, R. Rajesh and C. Tracy. I also thank the 
organizers
J.-P. Bouchaud and M. M\'ezard and all other participants of this summer school for physics, for fun,
and for making the school a memorable one.
        
%
%

%
\end{document}